\RequirePackage{fix-cm}
\documentclass[smallextended]{svjour3}       
\smartqed  
\usepackage{graphicx}

\usepackage{mathrsfs}

\usepackage{amsmath}
\usepackage{amssymb}
\usepackage{placeins}
 \usepackage{multirow}

\newcommand{\ket}[1]{|#1\rangle}

\begin{document}

\title{Quantum communication protocols as a benchmark for programmable quantum computers
\thanks{E. O. K. was supported by RFBR (project no. 18-37-00096). W. V. P. acknowledges a support from RFBR (project no. 15-02-02128). Yu. E. L. acknowledges a support from RFBR (project no. 17-02-01134) and the Program of Basic Research of HSE.}
}

\titlerunning{Quantum communication protocols as a benchmark for quantum computers}        

\author{A.\,A.~Zhukov         \and
    E.\,O.~Kiktenko \and
    A.\,A.~Elistratov \and
    W.\,V.~Pogosov \and
    Yu.\,E.~Lozovik 
}


\institute{A.\,A.~Zhukov \at
              Dukhov Research Institute of Automatics (VNIIA), 127055 Moscow, Russia \\
              National Research Nuclear University (MEPhI), 115409 Moscow, Russia
           \and
    E.\,O.~Kiktenko  \at
    Russian Quantum Center, Skolkovo, Moscow 143025, Russia \\
    Steklov Mathematical Institute of Russian Academy of Sciences, Moscow 119991, Russia \\
    Geoelectromagnetic Research Center of Schmidt Institute of Physics of the Earth, Russian Academy of Sciences, Troitsk, Moscow, 142190 Russia
    \and
    A.\,A.~Elistratov \at
    Dukhov Research Institute of Automatics (VNIIA), 127055 Moscow, Russia
    \and
    W.\,V.~Pogosov   \at
    Dukhov Research Institute of Automatics (VNIIA), 127055 Moscow, Russia \\
    Institute for Theoretical and Applied Electrodynamics, Russian Academy of
    Sciences, 125412 Moscow, Russia\\
    Moscow Institute of Physics and Technology, Dolgoprudny, Moscow Region 141700, Russia\\
    Tel.: +7(926)359-6034\\
    Fax.: +7(499)978-0903\\
    \email{Walter.Pogosov@gmail.com}           
    \and
    Yu.\,E.~Lozovik \at
    Dukhov Research Institute of Automatics (VNIIA), 127055 Moscow, Russia \\
    Institute of Spectroscopy, Russian Academy of Sciences, 142190 Moscow, Russia\\
    Moscow Institute of Electronics and Mathematics, National Research University Higher School of Economics, 101000 Moscow, Russia
}

\date{Received: date / Accepted: date}

\maketitle

\begin{abstract}
We point out that realization of quantum communication protocols in programmable quantum computers  provides a deep benchmark for capabilities of real quantum hardware. Particularly, it is prospective to focus on measurements of entropy-based characteristics of the performance and to explore whether a "quantum regime" is preserved. We perform proof-of-principle implementations of superdense coding and quantum key distribution BB84 using 5- and 16-qubit superconducting quantum processors of IBM Quantum Experience. We focus on the ability of these quantum machines to provide an efficient transfer of information between distant parts of the processors by placing Alice and Bob at different qubits of the devices. We also examine the ability of quantum devices to serve as quantum memory and to store entangled states used in quantum communication. Another issue we address is an error mitigation. Although it is at odds with benchmarking, this problem is nevertheless of importance in a general context of quantum computation with noisy quantum devices. We perform such a mitigation and noticeably improve some results.


\end{abstract}

\section{Introduction}

Quantum technologies based on manipulation with individual quantum objects and their quantum states are of great interest in problems of information transfer and processing. Historically, one of the first applications of quantum information technologies was quantum communication, and particularly, quantum key distribution (QKD)~\cite{BB84}.
This technology utilizes features of quantum light in order to provide information-theoretic (unconditional) security for classical data transmission and storage~\cite{Gisin2002,Scarani2009}.
Nowadays, there is a significant progress in experiments for providing long-distance point-to-point QKD links~\cite{Korzh2015,Frohlich2017,spaceQKD}, as well as establishing multi-site quantum networks~\cite{Elliott2005,Peev2009,Duplinskiy2017,Tysowski2018}.
There are also other important developing areas of quantum communication such as secure direct communication~\cite{Long2002,Hu2016,Zhang2017}, based on use of superdense coding~\cite{Bennet1992}, and transfering quantum states with quantum teleportation~\cite{Ren2017,Bennett1993}.

Another field, which is undergoing dramatic progress, is quantum computation with different physical platforms, among which superconducting quantum circuits as well as trapped ions seem to be most prospective, see, e.g., Refs. \cite{Girvin,Blatt}. Various quantum algorithms have been implemented to show concepts of error correction \cite{Martinis1,DiCarlo,Gambetta,Chow,Behera}, modeling spectra of molecules \cite{variat} and other fermionic systems \cite{Hubbard}, simulation of light-matter systems \cite{LM}, spin systems \cite{Martinisadiab}, many-body localization \cite{MBL}, machine learning \cite{ML}, scaling issues \cite{Rigetti} etc. However, in order to realize algorithms which are of practical importance, the quantum hardware must be characterized by the error rate, which is much lower than the error rate of state-of-the-art processors. In principle, fault-tolerant quantum computing can be achieved with the quantum error correction codes, but this strategy implies enormous overhead of quantum resources. As an alternative, different hybrid quantum-classical computation schemes supplemented by error mitigation approaches have been suggested with the hope that at least some practical implementations with near-term hardware can be achieved without a full error correction, see, e.g., Refs. \cite{variat,mitig1,mitig2,mitig3,mitig4,weare}.

In this paper, we combine two areas of quantum technologies -- quantum communication protocols and quantum computation. Our major idea is that quantum communication protocols can serve as deep benchmarks for capacity of programmable quantum computers. These quantum machines can be based on different physical realizations and characterized by different levels of connectivity. The quantum communication protocols rely on  ``quantum regime'' (or ``quantum advantage'') and it is of interest to examine how it survives in real quantum machines. Mathematically, ``quantum regime'' can be revealed through entropy-based quantities, such as mutual information or secret key length. The evaluation of such quantities in simulations with real quantum devices is an essential ingredient of our approach.

Let us mention that various benchmarks are also routinely used for classical computers. There are such tests which examine particular operations, while other benchmarks estimate a performance of a classical computer as a whole in connections to some particular classes of tasks. Our quantum benchmarks are similar to the tests of the second kind. We stress that such popular characteristics of qubits as their coherence times do not directly refer to the usability of a processor, which accumulates many other characteristics and also depends on a particular algorithm. Therefore, a development of various benchmarks seems to be an important and timely task. Notice that it was recently proposed in Ref. \cite{qvolume} to use a special metric termed as quantum volume for the power of quantum computers. The problem of benchmarking of quantum circuits was also discussed in Refs. \cite{referee1,referee2,referee3,referee4} either in the context of random circuits or particular quantum algorithms.

We here present the results of our proof-of-principle benchmarking experiments using superconducting quantum processors of IBM Quantum Experience available through the cloud service. Particularly, we used 5- and 16-qubit machines, the latter being considered as a rather complex quantum network with physical qubits modeling its nodes. We focused on the superdense coding protocol~\cite{Bennet1992} as well as on famous quantum key distribution BB84 protocol~\cite{BB84}. In contrast to earlier implementations of superdense coding in superconducting quantum computers, see, e.g., Ref. \cite{Sierra},  in our simulations we address entropy-based quantities and study an impact of physical device imperfections on the ability to transfer quantum information between different nodes of the quantum network by positioning Alice and Bob at different physical qubits. Such operations are of crucial importance, for example, in connection to simulations of nonlocal quantum models (including fermions or spins with long-range interactions) using quantum computers of limited connectivity. Another aspect we address is a capability of quantum processors to serve as a quantum memory for storing entangled states used in processes of quantum communication.

The paper is organized as follows.
In Sec.~\ref{sec:superdense} we consider a simulation of superdense coding.
In Sec.~\ref{sec:QKD} we analyze a performance of QKD with BB84 protocol.
We summarize our results in Sec.~\ref{sec:concl}.

\section{Quantum information transfer, quantum memory, and superdense coding}\label{sec:superdense}

\begin{figure}[h]
\center
    \includegraphics[width=0.95\linewidth]{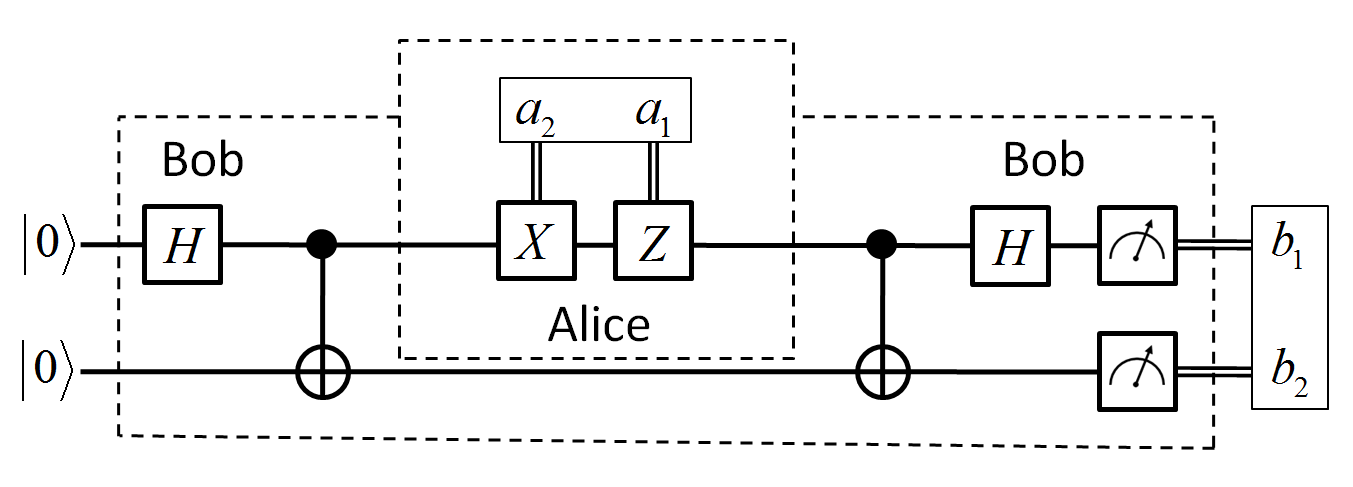}
    \caption{
        \label{sd}
        Schematic view of the simplest version of the superdense coding protocol.}
\end{figure}

Superdense coding is a quantum protocol, which allows increasing the information content using such a key resource of quantum systems as entanglement. The central idea is that two bits of classical information can be transferred with a single qubit participating in communication. Let us imagine that Alice intends to send two bits of information to Bob using qubits. Without relying on entanglement, Alice has to send two qubits in order to transfer two bits of information. However, the advantage appears provided Bob prepares two entangled qubits and sends only one of them to Alice. Alice encodes information in this qubit by applying two single-qubit gates and this qubit is then sent back to Bob. Bob performs Bell measurements of both qubits and extracts two bits of information despite of the fact that only single qubit has been utilized in quantum communication.

The whole scheme is illustrated in Fig. \ref{sd}.
Firstly, the entangled state is prepared using Hadamard gate
\begin{equation}
H = \frac{1}{\sqrt{2}}\begin{bmatrix}
1 & 1 \\
1 & -1 \\
\end{bmatrix},
\end{equation}
and controlled-NOT (CNOT) gate
\begin{equation}
	{\rm CNOT} = \begin{bmatrix}
		1 & 0 & 0 & 0 \\
		0 & 1 & 0 & 0 \\
		0 & 0 & 0 & 1 \\
		0 & 0 & 1 & 0 \\
	\end{bmatrix}
\end{equation}
(here the first and second qubits being controlling and target correspondingly).
Then,
Alice encodes two bits of information, $a_1$ and $a_2$, into a combination of two single-qubit gates: 00, 10, 01, and 11 are encoded into $II$, $ZI$, $IX$, and $ZX$, respectively.
Here $I$, $X$, $Y$ and $Z$ are standard Pauli (with identity) gates:
\begin{equation}
\begin{aligned}
&I=\begin{bmatrix}
1&0\\0&1
\end{bmatrix},\quad
&X=\begin{bmatrix}
0&1\\1&0
\end{bmatrix}, \\
&Y=\begin{bmatrix}
0&-i\\i&0
\end{bmatrix}, \quad
&Z=\begin{bmatrix}
1&0\\0&-1
\end{bmatrix}.
\end{aligned}
\end{equation}
These single-qubit gates are applied to the qubit sent by Bob to Alice, while by performing final Bell measurement Bob extracts two bits $b_1 b_2$.

In the absence of noises we have the identities $a_i=b_i$.
However, in real experimental realizations, due to the errors, the Bob's output value behaves like a random variable yet highly correlated but different from the Alice's input.
The quantity we are interested in is amount of information
\begin{equation}
{\mathcal I}(A,B) = H(B) - H(B|A),
\end{equation}
between the Alice's input $A=(a_1, a_2)$ and Bob's output $B=(b_1,b_2)$.
Here
\begin{equation}
H(X) = -\sum_{x}{\rm Pr}(X=x)\log_2{\rm Pr}(X=x)
\end{equation}
is a Shannon entropy of a random variable $X$ with possible values $\{x\}$ and
\begin{multline}
H(X|Y) = -\sum_{y}{\rm Pr}(Y=y)\times\\\sum_{x}{\rm Pr}(X=x|Y=y)\log_2{\rm Pr}(X=x|Y=y)
\end{multline}
is conditional entropy of $X$ given random variable $Y$ with possible values $\{y\}$.

It is easy to prove that for the uniform distribution of $A$ and absence of noises $\mathcal{I}(A,B)=2$ that corresponds to an ideal transfer of two bits of information.
The presence of errors decreases this value and condition $\mathcal{I}(A,B)\leq1$ implies
reaching a ``classical regime'' of information transfer.

In the next subsection, we model two real practical situations using both 5-qubit and 16-qubit quantum computers (IBMqx4 and IBMqx5, respectively): the process of information transfer from Alice to Bob positioned in distant nodes of the quantum network and impact of quantum memory imperfection for storing initial Bell states.

\subsection{Simulation of the information transfer through network nodes}

\begin{figure}[h]
\includegraphics[width=1.00\linewidth]{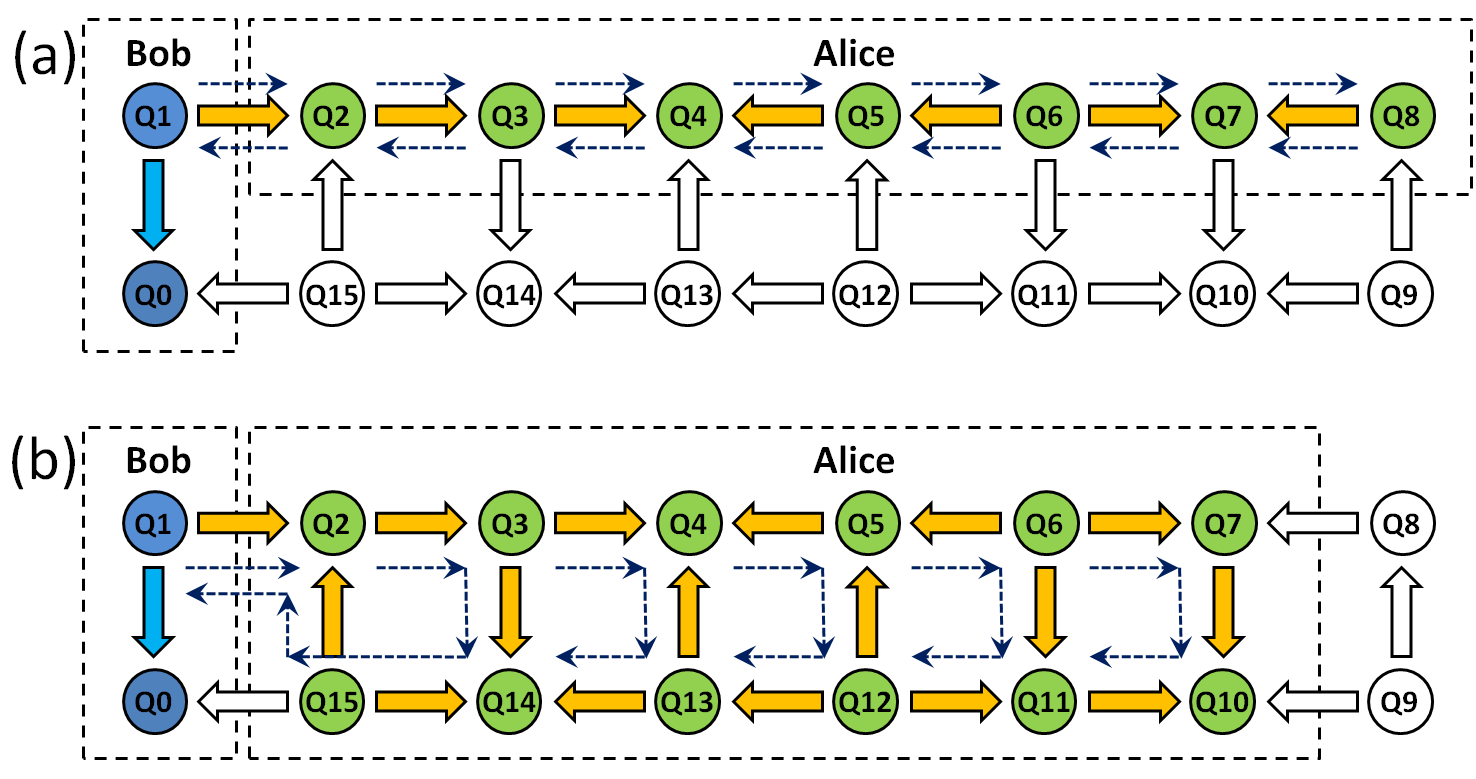}
\caption{
\label{IBMqx5}
Superdense coding implemented in 16-qubit IBMqx5 quantum processor. The transfer of information is localized in the upper row of the device (a) or within both rows of the device (b), as shown by dashed arrows. Big arrows indicate two-qubit gates and their directions.}
\end{figure}

We use 16-qubit processor to simulate information transfer through the quantum network nodes. Two different situations are analyzed, which are illustrated by  Fig.~\ref{IBMqx5}. Let us imagine that Bob initially has two entangled qubits, which are shown in Fig.~\ref{IBMqx5} as Q0 and Q1. Alice controls a single qubit located somewhere else in the upper row of qubits of the chip in the situation illustrated in Fig. \ref{IBMqx5} (a) or a single qubit, which can be located in the lower row either, see Fig. \ref{IBMqx5} (b). Bob sends the quantum state of its qubit Q1 to Alice by using SWAP operations between neighboring qubits in the upper row. Each SWAP can be composed from three CNOT gates, which is costly from the viewpoint of errors, since a typical error of each CNOT is currently several percent. In contrast, the errors of single-qubit gates are order of magnitude smaller.

After receiving the quantum state, which is already affected by imperfections of the device, Alice applies a couple of single-qubit gates and then sends the resultant quantum state back to Bob along either the same path within the upper row or through the lower row. Thus we now deal with the effects arising due to the imperfect quantum state transfer from one to another physical qubit. The most important characteristics, which provides a dominant contribution to the reduction of mutual information from the maximum value 2, is the number of SWAPs. For the purposes of our simulation we performed a set of four experiment with equal number of runs for four different values of $A$.
For each input $A$ we computed an entropy of output distribution $B$, and then obtained a final value of mutual information $\mathcal{I}(A,B)$.

\begin{figure}[h]
\includegraphics[width=1.00\linewidth]{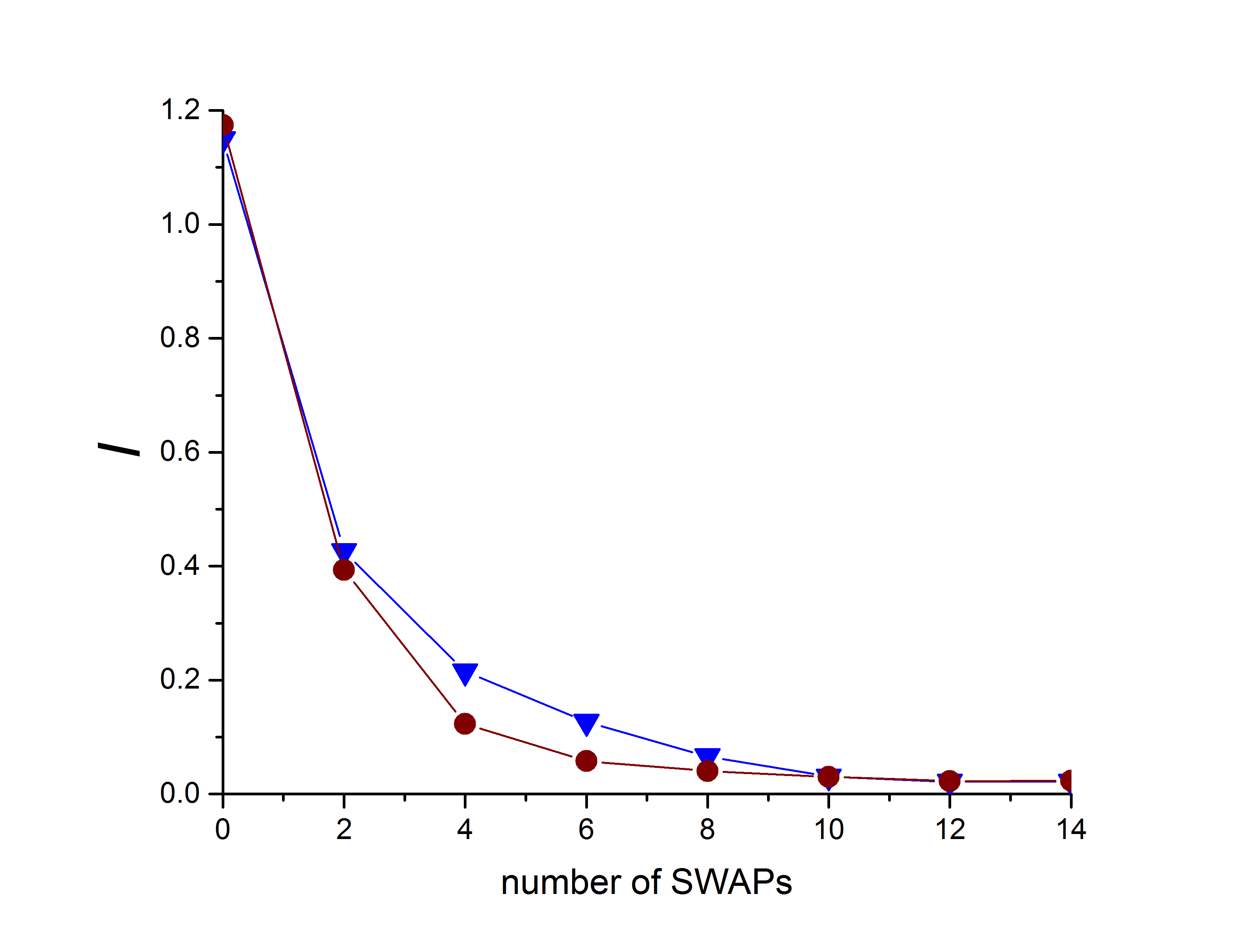}
\caption{
\label{SWAPS16q}
Superdense coding in 16-qubit device. Mutual information is plotted as a function of the number of SWAPs between Alice and Bob within the physical device. Two different sets of paths are considered, which are shown in Fig. \ref{IBMqx5} (a)  (blue triangles) and \ref{IBMqx5} (b) (brown circles).}
\end{figure}

Our results for the mutual information as a function of number of SWAPs are shown in Fig. \ref{SWAPS16q} by the blue circles in the situation depicted by Fig. \ref{IBMqx5} (a) and by brown triangles in the situation shown by Fig. \ref{IBMqx5} (b). Hereafter each given $(a_1 a_2)$ corresponds to 8192 shots (individual runs of the algorithm). For the illustration purposes, we also present raw data (output distributions) in the Appendix A for  Fig. \ref{IBMqx5} (a). It is seen that mutual information drops well below 1 just after two SWAPs. This configuration corresponds to the location of Alice at the nearest neighboring site to Bob. Mutual information becomes much smaller than 1 for long trajectories which involve multiple physical qubits of the chip. This result evidences that physical imperfections of the device are still too significant to provide an efficient information transfer from one part of the processor to another one.

\subsection{Simulation of the imperfections of quantum memory}

We now simulate the effect of imperfections of quantum memory, which stores entangled states used in superdense coding protocol. This imperfection is modeled by us by applying multiple identity gates $I$. Identity gates provide a time delay $\tau$, which leads to the decay of quantum states of physical qubits due to their interaction with the environment. The duration of an identity gate in IBMqx5 is $\tau=90$ nanoseconds, while both $T_1$ and $T_2$ are nearly 30-50 microseconds. The train of identity gates is applied before Alice makes encoding of classical information into her qubit.

\begin{figure}[h]
\includegraphics[width=.4\linewidth]{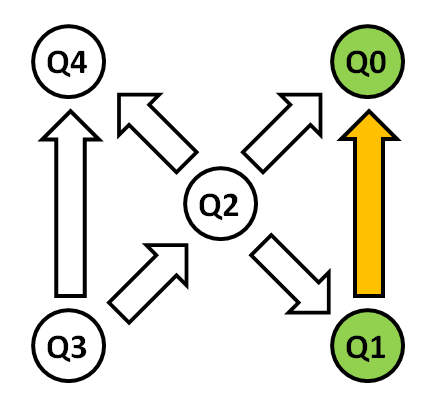}
\caption{
\label{IBMqx42}
Schematic image of IBMqx4 chip. Qubits Q0 and Q1 are utilized in our implementation of superdense coding protocol aimed to simulate the effect of decoherence in the quantum memory. Arrows indicate two-qubit gates and their directions.}
\end{figure}

\begin{figure}[h]
\includegraphics[width=1.00\linewidth]{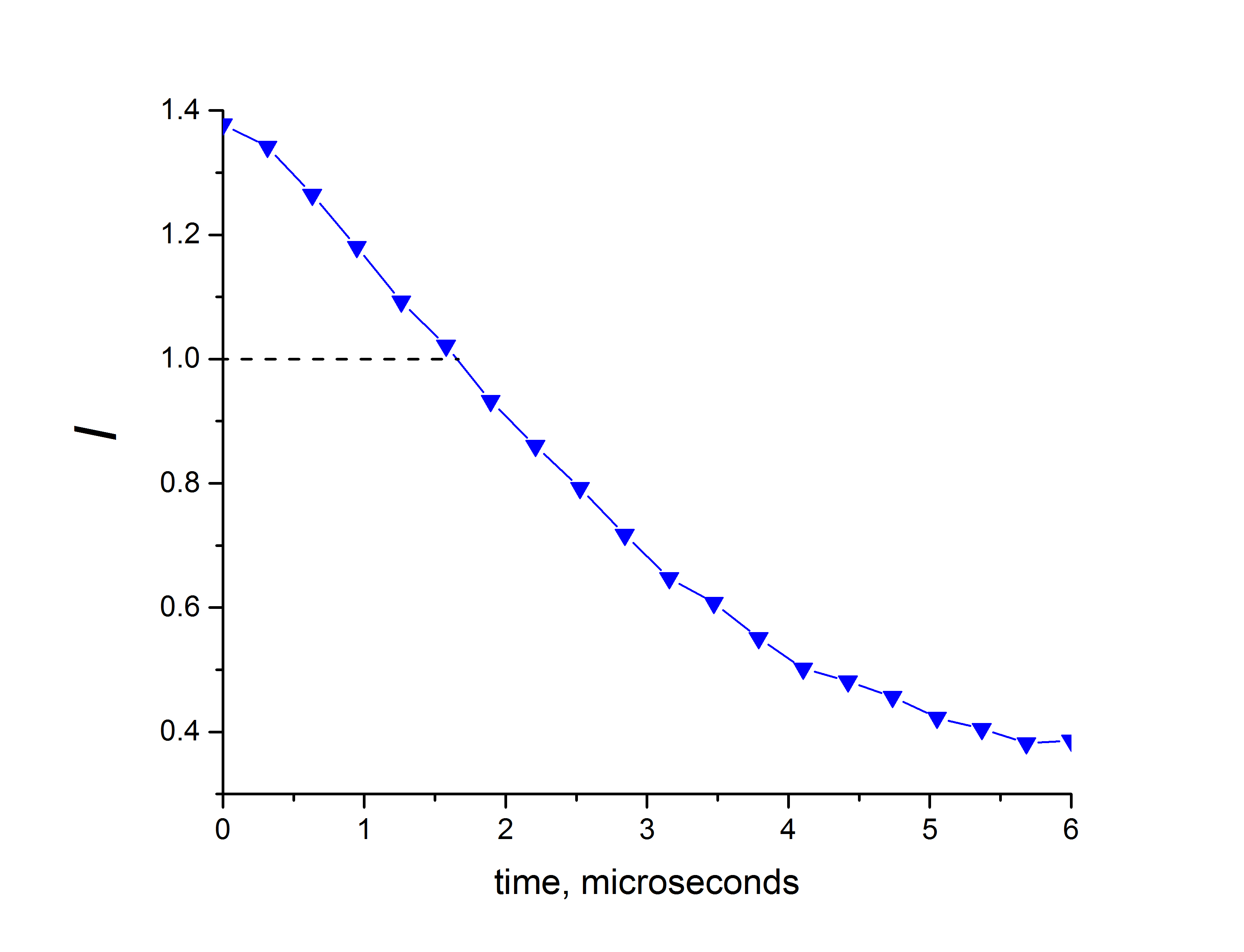}
\caption{
\label{sd5q}
Superdense coding in 5-qubit device. Mutual information is shown as a function of the storage time.}
\end{figure}

In our simulations, we first use 5-qubit IBMqx4 processor, which is shown schematically in Fig. \ref{IBMqx42}. Two physical qubits Q0 and Q1 of the device are utilized. Figure \ref{sd5q} shows the result of our simulation of mutual information as a function of the time to store entangled states. The output distributions are presented in Appendix A. We see that even in the absence of identity gates the observed value of mutual information is less than 2. The major reasons are the imperfections of CNOT gates and existing readout errors, the latter being typically several percent. Anyway, we see a ``quantum advantage'' up to a certain number of identity gates applied. This fact is very important, since it evidences that superdense coding is indeed realized in this real device -- more than a single bit of information is transmitted. The mutual information decays exponentially from the initial value. It becomes lower than 1 after nearly 2 microseconds, which implies that the effect of imperfection of "quantum memory" becomes too high to support superdense coding.

Note that there is a connection between our simulations and the approach of Ref. \cite{mitig1}, where it was proposed to intentionally enhance errors by inserting additional gates in the quantum circuit in such a way that they are reduced to the identity gate for an ideal system (for example, a couple of CNOTs gives the identity gate); then the extrapolated zero-error quantities can be extracted from noisy experimental data. This trick was recently used in cloud computing of atomic nucleus \cite{nucleus}.

\begin{figure}[h]
\includegraphics[width=1.00\linewidth]{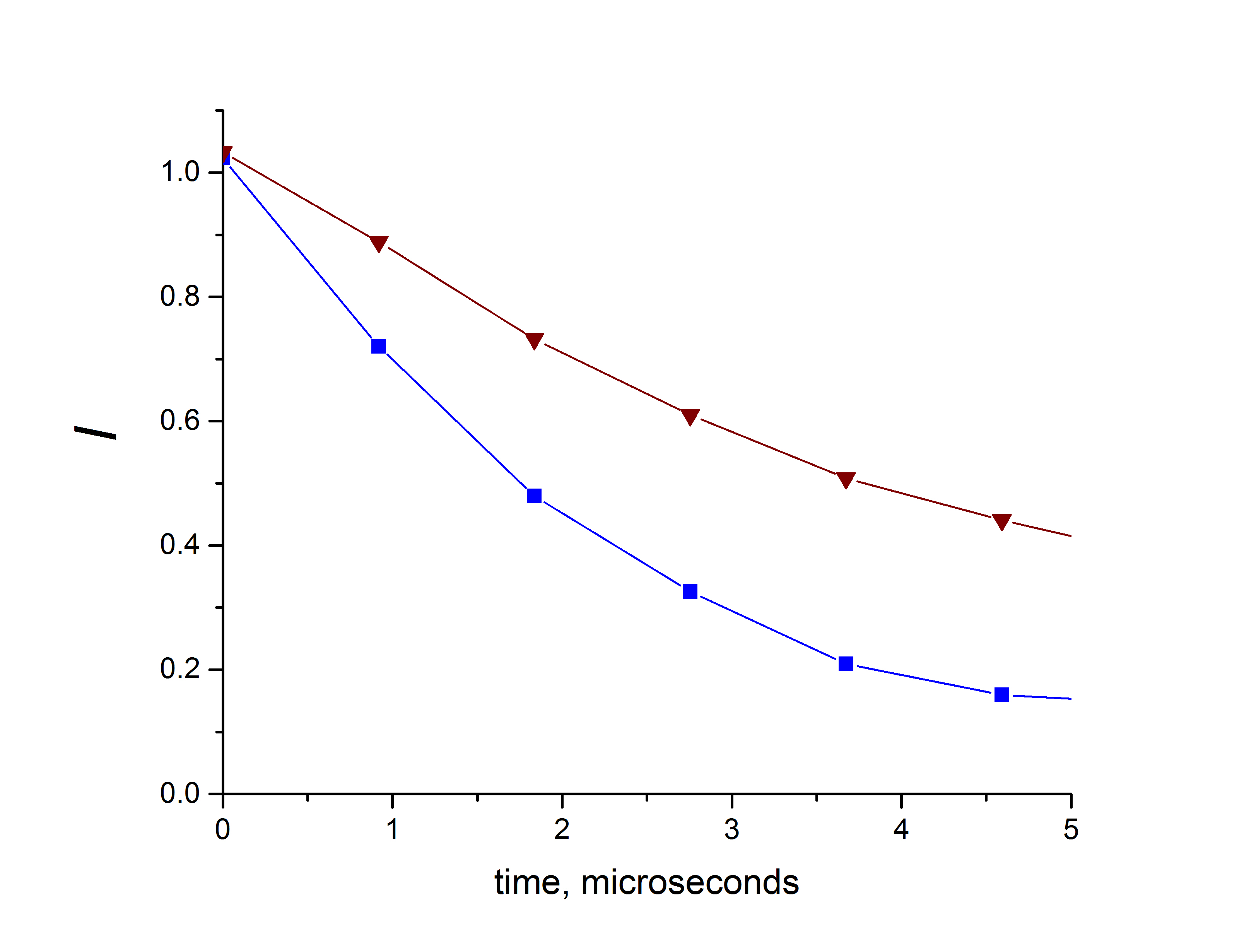}
\caption{
\label{sd16q}
Superdense coding in 16-qubit device. Mutual information is shown as a function of the storage time with the correction of coherent errors (upper curve) and without it (lower curve).}
\end{figure}

Now we implement the same algorithm using 16-qubit IBMqx5 chip having in mind to explore the ability of this larger chip to serve as a "quantum memory" used to store quantum states. Qubits Q0 and Q1 are utilized in the simulation. The results of our simulations are shown in Fig. \ref{sd16q} by blue squares, while Appendix A provides the output distributions. The main difference with the IBMqx4 chip is in the reduction of the initial value of the mutual information, which becomes only slightly larger than 1. The reason is in the slightly enhanced errors of two-qubit gates as well as readouts. We again stress that the mutual information is very sensitive to physical errors. However, we would like also to mention that the results are not completely stable and can vary in time after each calibration of the device.

We also performed an additional analysis of physical imperfections of IBMqx5 performance, see Appendix B for the details. We found that one of the major contributions to errors relevant for our simulations is provided by specific processes, which systematically change internal phases in entangled states. For example, they lead to oscillations between Bell states $\ket{\Psi_+}$ and $\ket{\Psi_-}$ on the time scale $t_{\rm osc} \simeq 10$ microseconds. We managed to partially correct these coherent errors by applying the single-qubit gate
\begin{equation}
	U(\varphi)=\begin{bmatrix}
	1 & 0 \\
	0 & e^{i\varphi}
	\end{bmatrix}
\end{equation}
after the train of identity gates, where $\varphi=-\pi t / t_{\rm osc}$.

The corrected results are shown in Fig. \ref{sd16q} by brown triangles, while the output distributions are given in Appendix A. Although the dependence of mutual information on time of course starts from the same value at zero time, the decay of mutual information is significantly slowed down. Note that the procedure we use applies to a particular quantum processor and quantum algorithm. Of course, error mitigation somehow contradicts our central idea of quantum benchmarking, but this aspect is of general importance in the context of quantum computation with noisy quantum hardware.

Notice that recently a well known dynamically decoupling approach was applied for superconducting quantum computers to protect qubits from the environment \cite{DD}.

\section{Quantum key distribution with superconducting processor}\label{sec:QKD}

In this Section, we use 5-qubit chip to implement well known protocol BB84.

We start with a brief description of BB84 protocol.
Its aim is to generate two identical random bit strings (keys) on both communicating sides (Alice and Bob) is such a way that these strings are only known by Alice and Bob.
The work flow of the protocol is the following.
\begin{enumerate}
  \item Alice generates a random $L$-bit string $K_A^{\rm raw}\in\{0,1\}^L$, which corresponds to key bit.
  \item Alice generates a random $L$-bit string $B_A^{\rm raw}\in\{+,\times\}^L$, where special symbols $+$ and $\times$ correspond to choice of basis for encoding a bit.
  \item For each pair $(K_A^{\rm raw}[i],B_A^{\rm raw}[i])$ Alice prepares a single-photon state via the following mapping
      \begin{equation}
        \begin{aligned}
          &(0,+) \rightarrow \ket{H},\\
          &(1,+) \rightarrow \ket{V},\\
          &(0,\times) \rightarrow \ket{D}=\frac{1}{\sqrt{2}}(\ket{H}+\ket{V}),\\
          &(1,\times) \rightarrow \ket{A}=\frac{1}{\sqrt{2}}(\ket{H}-\ket{V}),
        \end{aligned}
      \end{equation}
      where $\ket{H}$, $\ket{V}$, $\ket{D}$ and $\ket{A}$ are horizontally, vertically, diagonally and anti-diagonally polarized states of photons.
  \item All the photons are transferred to Bob.
  \item Bob performs its polarization measurement in one of two randomly chosen bases: $\ket{H}/\ket{V}$ (``$+$'') or $\ket{D}/\ket{A}$ (``$\times$''), and stores results of the measurement in bit string $K_B^{\rm raw}$.
  \item After all photons were measured (note, that in real experiments there are losses, and not all the photons sent by Alice reach Bob), Alice and Bob exchange with information about the bases they used in preparation and measurement using classical authenticated channel.
      Then they discard all the bits from $K_A^{\rm raw}$ and $K_B^{\rm raw}$, which corresponds to events of preparation and measurement in inconsistent bases.
      The resulting bit strings $K_A^{\rm sift}$ and $K_B^{\rm sift}$ are named sifted keys.
  \item The sifted keys are input for a classical post-processing procedure, which consists of (i) parameter estimation, aimed on evaluating an error $q$ between $K_A^{\rm sift}$ and $K_B^{\rm sift}$ (named quantum bit error rate QBER), (ii) information reconciliation, aimed on making the keys of Alice and Bob to be identical, and finally (iii) privacy amplification, which aimed on removing partial information of an eavesdropper about reconciled keys.
      Neglecting final-length effects, the size of the resulting (identical and secure) keys $K_A^{\rm sift}$ and $K_B^{\rm sift}$ is given by
      \begin{equation} \label{seckeylength}
        l_{\rm sec} = N(1-h(q))-Nf_{\rm ec}h(q),
      \end{equation}
      where $N$ is length of sifted keys,
      \begin{equation}
        h(q) = -q\log_2q-(1-q)\log_2(1-q)
      \end{equation}
      is binary entropy function and $f_{\rm ec}$ is ``efficiency'' of information reconciliation algorithm (in all the further considerations we take $f_{\rm ec}=1.15$, that correspond to real practise~\cite{Kiktenko2017}).
      The expression~\eqref{seckeylength} gives a length to which the reconciled sifted keys should be shortened by employing publicly announced random hash function from universal$_2$ set at the stage of privacy amplification~\cite{Tomamichel2012}.
      Note, that negatives values of $l_{\rm sec}$ correspond to the fact of impossibility to distill the provably secure keys.
\end{enumerate}

\subsection{Simulation of the decoherence in quantum memory}

\begin{figure}[h]
\includegraphics[width=.95\linewidth]{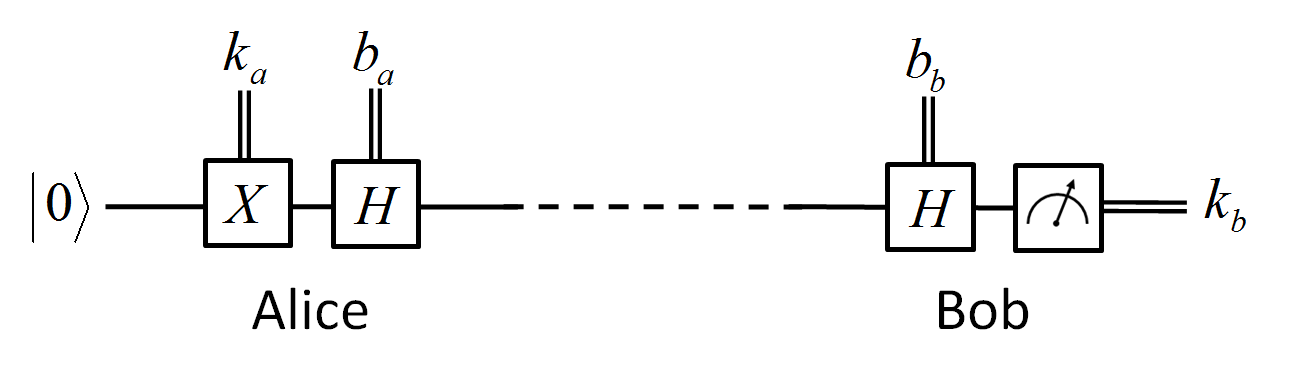}
\caption{
\label{bb}
Illustration of BB84 protocol implemented with a single physical qubit of IBMx4 chip to simulate the effect of imperfections of quantum memory used to store entangled states.}
\end{figure}

Our first experiment deals with a single physical qubit, which is Q1 in the schematic image of the processor IBMqx4, see Fig. \ref{IBMqx42}. Alice encodes 0 or 1 of a key in this qubit using single-qubit gates $I$ or $X$, respectively. After that, we choose the basis ``+'' or ``$\times$'' by applying single-qubit gates $I$ or Hadamard gate $H$
respectively. Then, we apply a train of identity gates to simulate a time delay in the transmission line between Alice and Bob. Finally, Bob, which is physically located at the same site within our simulation, measures this qubit in the same basis ``$+$'' or ``$\times$''. In order to reduce the number of experiments, we also assume that the basis of Alice and the basis of Bob are the same, thus we analyze a sifted key. The whole protocol is illustrated in Fig. \ref{bb}.

\begin{figure}[h]
\includegraphics[width=1.00\linewidth]{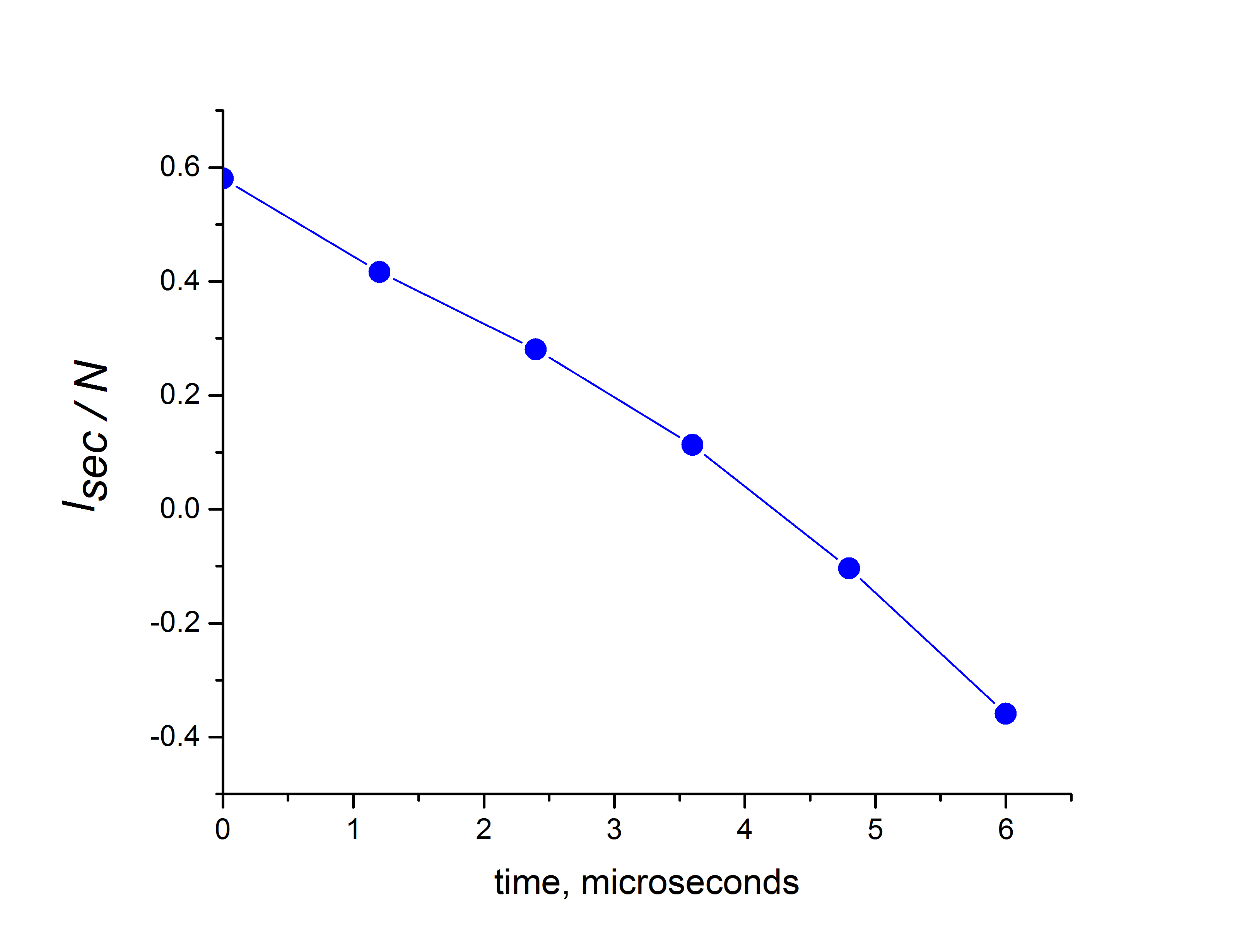}
\caption{
\label{BBdelay}
Experimentally determined secure key length $l_{\rm sec}$ as a function of the delay time.}
\end{figure}

Using Eq. \eqref{seckeylength}, we find $l_{\rm sec}$. Fig. \ref{BBdelay} shows $l_{\rm sec}$ as a function of the delay time. It is seen from this figure that $l_{\rm sec}$ vanishes after nearly four microseconds. This result is similar to the above result for the mutual information within the protocol of the superdense coding. The ``decay time'' is several microseconds, which is nearly one order of magnitude smaller than $T_1$ and $T_2$ of physical qubits. This illustrates certain limitations of $T_1$ and $T_2$ as unambiguous characteristics of quantum processors. The error distributions for this experiment are given in Appendix C.

\subsection{Simulation of the information transfer through network nodes}

In order to simulate information transfer between network nodes, we use 5-qubit IBMqx4 chip because of the lower error rate. We have chosen qubits Q0 and Q1 of the device for our simulation because physical errors associated with these particular qubits are also minimum. This is of particular importance in the view of sensitivity of entropy-like characteristics to errors.

\begin{figure}[h]
\includegraphics[width=.95\linewidth]{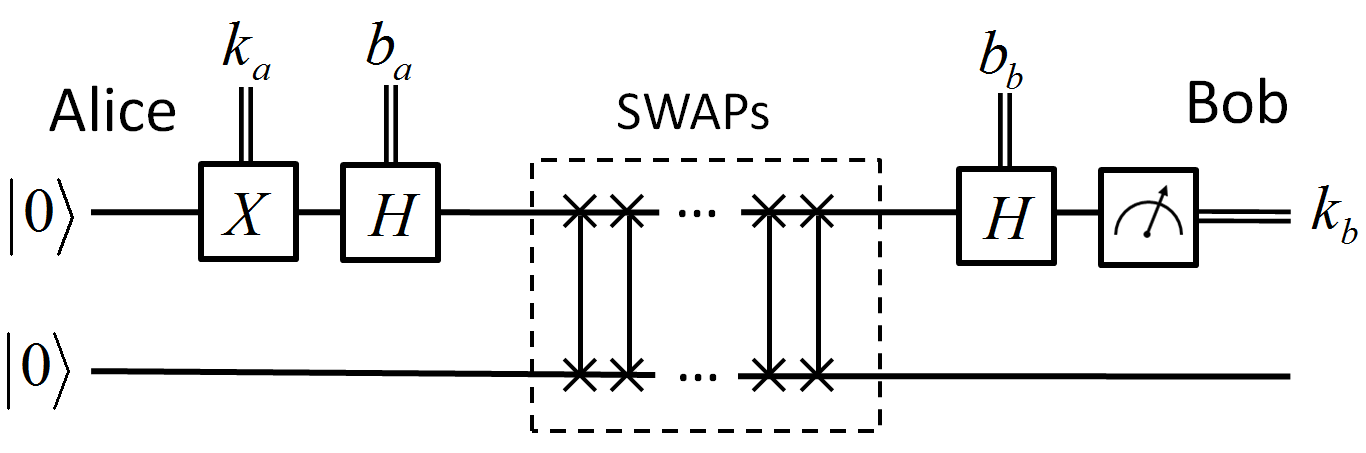}
\caption{
\label{bb1}
Illustration of BB84 protocol implemented using a single physical qubit of IBMx4 chip as a logical qubit.}
\end{figure}

Let us imagine that both Alice and Bob are situated at qubit Q0 of the device. Alice again encodes 0 or 1 and chooses the basis. We model an information transfer by performing a sequence of SWAP gates between Q0 and Q1. After the even number of SWAPs the quantum state, influenced by imperfections of the device, returns to Q0. After receiving the resultant quantum state, Bob makes the same manipulations as in the previous simulation (see the preceding subsection). The whole protocol is illustrated in Fig. \ref{bb1}.

The results of our simulations are shown in Fig. \ref{BBswaps} by blue circles. The error distributions are given in Appendix C. It is seen from this figure that $l_{\rm sec}/N$ remains positive until four SWAPs. Positiveness of this quantity is a necessary condition to establish a secure communication. Let us mention that $l_{\rm sec}/N$ is more robust with respect to the number of SWAPs compared to the mutual information in the protocol of the superdense coding. This is due to the fact that the latter requires two entangled qubits, so that CNOT errors play a more destructive role.

\begin{figure}[h]
\includegraphics[width=1.00\linewidth]{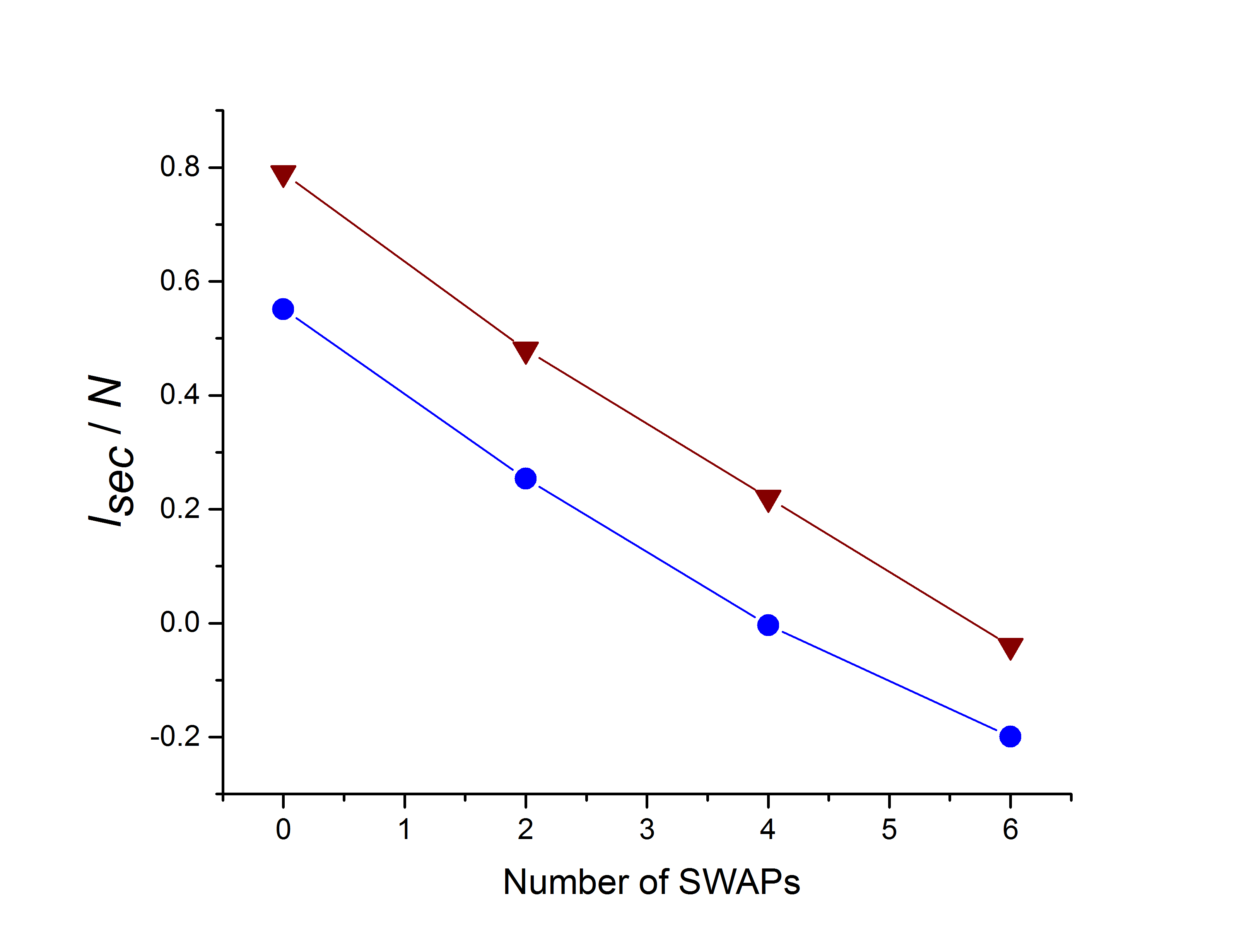}
\caption{
\label{BBswaps}
$l_{\rm sec}$ as a function of the number of SWAPs with encoding the single logical qubit into single physical qubit (blue circles) and into a couple of physical qubits supplemented by the post-selection of the results (brown triangles).}
\end{figure}

\begin{figure}[h]
\includegraphics[width=.95\linewidth]{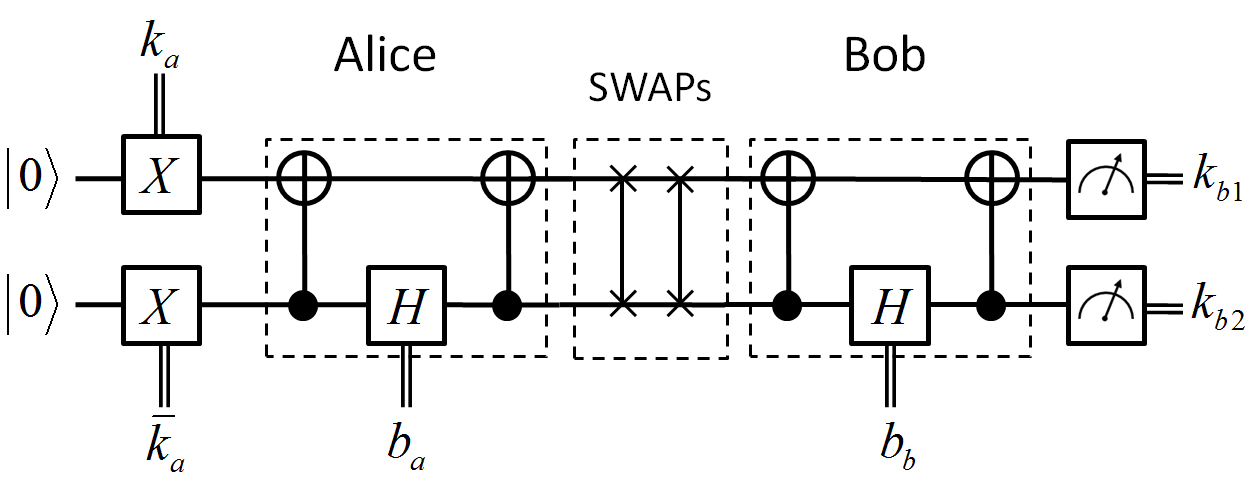}
\caption{
\label{bb2}
Illustration of BB84 protocol implemented using a couple of physical qubits of IBMx4 chip to encode single logical qubit.}
\end{figure}

An analysis shows that one of the major sources of errors in our simulation is the asymmetry of readout results: the probability to correctly measure excited state of the qubit is slightly lower than the probability to correctly measure ground state of the qubit. In order to mitigate this error, we encode single logical qubit into two physical qubits. We define states of the logical qubit as $|0 \rangle_{\rm logic}= | 10 \rangle$ and $| 1 \rangle_{\rm logic} = |01 \rangle$, and perform a following way of encoding quantum states
\begin{equation}
\begin{aligned}
&\ket{k_a, \overline{k}_a} &\text{ for } b_a=0, \\
&\frac{1}{\sqrt{2}}(\ket{01}+(-1)^{k_a}\ket{10}) &\text{ for } b_a=1,	
\end{aligned}
\end{equation}
where $\overline{k}_a := 1 -k_a$. After executing the algorithm, we discard the results of the form $| 00 \rangle$ and $| 11 \rangle$, which belong to another subspace of Hilbert space. This allows us to improve our results, despite of the fact that all single-qubit gates are now replaced by two-qubit gates. In addition, this approach makes our simulations closer to polarization encoding in quantum optics, where one-photon horizontally and vertically polarized states $\ket{H}$ and $\ket{V}$ might be considered as states $\ket{1}_H\ket{0}_V$ and $\ket{0}_H\ket{1}_V$ in Fock representation (here subscripts $H$ and $V$ correspond to horizontally and vertically polarized modes of the field).

For these simulations, we used qubits Q0 and Q1. The scheme is illustrated in Fig. \ref{bb2}. In this case, both Alice and Bob are located at the pair of physical qubits Q0 and Q1 at once. SWAPs were again performed between physical qubits, thus quantum states of these two qubits are interchanged upon each SWAP. In the case of an ideal system, the state of the logical qubit should not change under even number of SWAPs. The results of our simulations are shown in Fig. \ref{BBswaps} by brown triangles. Compared to the result without such encoding and the post-selection (blue circles), these new results appear to be much more robust with respect to the number of SWAPs. The error distributions are again given in Appendix C.

Notice that we also implemented a similar algorithm using 16-qubit device, where full SWAPs of logical states composed from the states of two physical qubits have been performed. In this case, post-selection procedure was less efficient than in the 5-qubit device, but the improvement has still been achieved. These data are not presented here.

\section{Conclusions}\label{sec:concl}

In this article, we suggested that programmable quantum processors can be utilized as a platform for implementation of quantum communication protocols: the realization of these protocols in existing and future quantum computers may provide deep benchmarks for their capabilities. An important ingredient of our suggestion is an experimental determination of mutual information or some other entropy-based quantity which rigorously quantifies an efficiency of the protocol realization. Thus, our approach has to be contrasted with the popular scheme of randomized benchmarking, see, e.g., Refs. \cite{Magesan,referee4,what} and has some similarities with other recent suggestions dealing with particular important algorithms \cite{referee2}. Let us stress that our ideas are quite universal in the sense that they can be applied to quantum computers based on different physical realizations and with different levels of connectivity.

We performed proof-of-principle simulations of two protocols using superconducting quantum computers of IBM Quantum Experience. In particular, we used 5- and 16-qubit superconducting processors. The latter can be considered as quite complex quantum network with physical qubits being nodes of the network. We concentrated on the protocol of superdense coding as well as famous quantum key distribution BB84 protocol. In our simulations, we mostly focused on quantum information transfer between different parts of the processors by placing Alice and Bob in separated nodes (physical qubits) and transferring quantum information between them using sequences of SWAP gates. Another issue we addressed was an ability of a quantum processor to serve as a "quantum memory" and to store entangled states used in quantum communication.

We found that the imperfections of the quantum machines we used were too significant to support an efficient quantum information transfer between distant qubits of the devices. In addition, the typical storage time of entangled states, which maintains "quantum regime" in our simulations, turns out to be much smaller than $T_1$ and $T_2$ of individual qubits.

Our experiments with noisy quantum machines also provide a playground for such an important activity as error mitigation although this issue, strictly speaking, is at odds with quantum benchmarking itself. Nevertheless, we suggested and applied certain tricks, which enabled us to mitigate and partially suppress errors of the devices. Namely, we used different types of qubit encoding supplemented by a proper post-selection as well as additional unitary rotations aimed to compensate undesirable but coherent phase drifts in Bell states. These tricks, which are dependent on particular algorithms and processors, lead to noticeable improvements of the results of our simulations.

We believe that our ideas will be useful both for in the context of quantum computation with quantum machines based on different physical platforms, for errors mitigation, as well as for studies of quantum communication protocols and quantum networks.

\textbf{Acknowledgments.} -- We acknowledge use of the IBM Quantum Experience for this
    work. The viewpoints expressed are those of the authors and
    do not reflect the official policy or position of IBM or the
    IBM Quantum Experience team. E. O. K. was supported by RFBR (project no. 18-37-00096). W. V. P. acknowledges a support from RFBR (project no. 15-02-02128). Yu. E. L. acknowledges a support from RFBR (project no. 17-02-01134) and the Program of Basic Research of HSE.

\appendix

\section{Output distributions for superdense coding}

Table 1 shows output distributions for the superdense coding protocol for the situation corresponding to  Fig. \ref{IBMqx5} (a) for different number of SWAPs, as obtained from 16-qubit IBMqx5 device. Here $(a_1, a_2)$ is Alice's input, while $(b_1,b_2)$ is Bob's output. Results presented in the table provide output distributions in connection to the input data. In the ideal situation, the input and output must be the same, so that the corresponding matrix for each given $(a_1, a_2)$ should be identity (unit) matrix. We see from Table 1 that, in reality, even for the zero number of SWAPs this matrix is rather different from the identity matrix.

\begin{table}[!hbp]
\centering
  \caption{The output distribution for superdense coding protocol for different number of SWAPs. For each input, 8192 runs of the algorithm on 16-qubit IBMqx5 device have been performed.}
\begin{tabular}{|c|c|c|c|c|c|}
\hline
\multirow{2}{*}{SWAPs} & \multirow{2}{*}{$a_1$, $a_2$} & \multicolumn{4}{c|}{$b_1$, $b_2$} \\ \cline{3-6}
                       &                               & 0,0    & 1,0    & 0,1    & 1,1    \\ \hline
\multirow{4}{*}{0}     & 0,0                           & 0.940  & 0.022  & 0.031  & 0.008  \\
                       & 1,0                           & 0.117  & 0.815  & 0.029  & 0.039  \\
                       & 0,1                           & 0.121  & 0.015  & 0.840  & 0.024  \\
                       & 1,1                           & 0.031  & 0.114  & 0.115  & 0.739  \\ \hline
\multirow{4}{*}{2}     & 0,0                           & 0.684  & 0.078  & 0.172  & 0.067  \\
                       & 1,0                           & 0.154  & 0.551  & 0.094  & 0.201  \\
                       & 0,1                           & 0.250  & 0.063  & 0.617  & 0.069  \\
                       & 1,1                           & 0.113  & 0.265  & 0.136  & 0.486  \\ \hline
\multirow{4}{*}{4}     & 0,0                           & 0.595  & 0.127  & 0.164  & 0.114  \\
                       & 1,0                           & 0.190  & 0.454  & 0.143  & 0.213  \\
                       & 0,1                           & 0.263  & 0.117  & 0.511  & 0.109  \\
                       & 1,1                           & 0.177  & 0.256  & 0.173  & 0.393  \\ \hline
\multirow{4}{*}{6}     & 0,0                           & 0.510  & 0.145  & 0.219  & 0.126  \\
                       & 1,0                           & 0.240  & 0.430  & 0.166  & 0.164  \\
                       & 0,1                           & 0.324  & 0.151  & 0.396  & 0.129  \\
                       & 1,1                           & 0.194  & 0.227  & 0.193  & 0.386  \\ \hline
\multirow{4}{*}{8}     & 0,0                           & 0.406  & 0.172  & 0.276  & 0.147  \\
                       & 1,0                           & 0.253  & 0.370  & 0.184  & 0.193  \\
                       & 0,1                           & 0.326  & 0.166  & 0.366  & 0.142  \\
                       & 1,1                           & 0.212  & 0.249  & 0.205  & 0.334  \\ \hline
\multirow{4}{*}{10}    & 0,0                           & 0.374  & 0.188  & 0.287  & 0.151  \\
                       & 1,0                           & 0.257  & 0.314  & 0.209  & 0.220  \\
                       & 0,1                           & 0.353  & 0.176  & 0.313  & 0.157  \\
                       & 1,1                           & 0.250  & 0.264  & 0.218  & 0.268  \\ \hline
\multirow{4}{*}{12}    & 0,0                           & 0.357  & 0.197  & 0.282  & 0.163  \\
                       & 1,0                           & 0.264  & 0.293  & 0.212  & 0.231  \\
                       & 0,1                           & 0.360  & 0.179  & 0.297  & 0.164  \\
                       & 1,1                           & 0.257  & 0.268  & 0.225  & 0.250  \\ \hline
\multirow{4}{*}{14}    & 0,0                           & 0.357  & 0.197  & 0.283  & 0.164  \\
                       & 1,0                           & 0.264  & 0.293  & 0.212  & 0.231  \\
                       & 0,1                           & 0.360  & 0.180  & 0.297  & 0.164  \\
                       & 1,1                           & 0.257  & 0.268  & 0.225  & 0.250  \\ \hline
\end{tabular}
\end{table}

Table 2 presents similar data for different values of delay time, as obtained from 5-qubit IBMqx4 device. We again see noticeable deviations from the ideal distribution even for zero waiting time.

\begin{table}[!hbp]
\centering
  \caption{The output distribution for superdense coding protocol for different values of time delay. For each input, 8192 runs of the algorithm on 5-qubit IBMqx4 device have been performed.}
\begin{tabular}{|c|c|c|c|c|c|}
\hline
\multirow{2}{*}{Time, $\mu s$} & \multirow{2}{*}{$a_1$, $a_2$} & \multicolumn{4}{c|}{$b_1$, $b_2$} \\ \cline{3-6}
                               &                               & 0,0    & 1,0    & 0,1    & 1,1    \\ \hline
\multirow{4}{*}{0.0}           & 0,0                           & 0.950  & 0.018  & 0.024  & 0.008  \\
                               & 1,0                           & 0.083  & 0.885  & 0.010  & 0.022  \\
                               & 0,1                           & 0.083  & 0.007  & 0.893  & 0.016  \\
                               & 1,1                           & 0.014  & 0.070  & 0.083  & 0.833  \\ \hline
\multirow{4}{*}{1.3}           & 0,0                           & 0.889  & 0.029  & 0.061  & 0.020  \\
                               & 1,0                           & 0.093  & 0.824  & 0.024  & 0.059  \\
                               & 0,1                           & 0.128  & 0.021  & 0.822  & 0.028  \\
                               & 1,1                           & 0.032  & 0.121  & 0.091  & 0.756  \\ \hline
\multirow{4}{*}{2.5}           & 0,0                           & 0.792  & 0.044  & 0.137  & 0.028  \\
                               & 1,0                           & 0.094  & 0.731  & 0.044  & 0.131  \\
                               & 0,1                           & 0.195  & 0.037  & 0.729  & 0.040  \\
                               & 1,1                           & 0.054  & 0.209  & 0.089  & 0.649  \\ \hline
\multirow{4}{*}{3.8}           & 0,0                           & 0.679  & 0.056  & 0.226  & 0.039  \\
                               & 1,0                           & 0.102  & 0.619  & 0.059  & 0.220  \\
                               & 0,1                           & 0.286  & 0.049  & 0.616  & 0.050  \\
                               & 1,1                           & 0.076  & 0.319  & 0.092  & 0.514  \\ \hline
\multirow{4}{*}{5.1}           & 0,0                           & 0.565  & 0.061  & 0.324  & 0.050  \\
                               & 1,0                           & 0.101  & 0.510  & 0.074  & 0.315  \\
                               & 0,1                           & 0.386  & 0.053  & 0.501  & 0.061  \\
                               & 1,1                           & 0.089  & 0.407  & 0.094  & 0.410  \\ \hline
\multirow{4}{*}{6.0}           & 0,0                           & 0.496  & 0.065  & 0.386  & 0.054  \\
                               & 1,0                           & 0.105  & 0.447  & 0.078  & 0.370  \\
                               & 0,1                           & 0.459  & 0.063  & 0.417  & 0.061  \\
                               & 1,1                           & 0.094  & 0.456  & 0.093  & 0.357  \\ \hline
\end{tabular}
\end{table}

Table 3 and Table 4 provide output distributions without error correction and with error correction, respectively, for different values of waiting time, as obtained from 16-qubit IBMQx5 device. We again see that the distributions are rather different from ideal ones even at $t=0$, but the error correction, in general, indeed improves the results.

\begin{table}[!hbp]
\centering
  \caption{The output distribution for superdense coding protocol for different values of time delay without any correction of coherent errors. For each input, 8192 runs of the algorithm on 16-qubit IBMqx5 device have been performed.}
\begin{tabular}{|c|c|c|c|c|c|}
\hline
\multirow{2}{*}{\begin{tabular}[c]{@{}c@{}}Time, $\mu s$\end{tabular}} & \multirow{2}{*}{$a_1$, $a_2$} & \multicolumn{4}{c|}{$b_1$, $b_2$} \\ \cline{3-6}
                                                                         &                               & 0,0    & 1,0    & 0,1    & 1,1    \\ \hline
\multirow{4}{*}{0.0}                                                     & 0,0                           & 0.945  & 0.011  & 0.043  & 0.001  \\
                                                                         & 1,0                           & 0.144  & 0.775  & 0.030  & 0.051  \\
                                                                         & 0,1                           & 0.156  & 0.026  & 0.765  & 0.053  \\
                                                                         & 1,1                           & 0.044  & 0.135  & 0.128  & 0.694  \\ \hline
\multirow{4}{*}{0.9}                                                     & 0,0                           & 0.794  & 0.090  & 0.074  & 0.042  \\
                                                                         & 1,0                           & 0.156  & 0.728  & 0.054  & 0.061  \\
                                                                         & 0,1                           & 0.163  & 0.057  & 0.706  & 0.074  \\
                                                                         & 1,1                           & 0.079  & 0.147  & 0.135  & 0.638  \\ \hline
\multirow{4}{*}{1.8}                                                     & 0,0                           & 0.699  & 0.117  & 0.118  & 0.066  \\
                                                                         & 1,0                           & 0.170  & 0.641  & 0.082  & 0.107  \\
                                                                         & 0,1                           & 0.204  & 0.084  & 0.617  & 0.095  \\
                                                                         & 1,1                           & 0.109  & 0.183  & 0.151  & 0.556  \\ \hline
\multirow{4}{*}{2.8}                                                     & 0,0                           & 0.620  & 0.118  & 0.179  & 0.082  \\
                                                                         & 1,0                           & 0.170  & 0.574  & 0.098  & 0.159  \\
                                                                         & 0,1                           & 0.269  & 0.101  & 0.528  & 0.102  \\
                                                                         & 1,1                           & 0.131  & 0.234  & 0.158  & 0.477  \\ \hline
\multirow{4}{*}{3.7}                                                     & 0,0                           & 0.531  & 0.129  & 0.244  & 0.096  \\
                                                                         & 1,0                           & 0.181  & 0.485  & 0.120  & 0.215  \\
                                                                         & 0,1                           & 0.339  & 0.112  & 0.438  & 0.110  \\
                                                                         & 1,1                           & 0.149  & 0.287  & 0.156  & 0.408  \\ \hline
\multirow{4}{*}{4.6}                                                     & 0,0                           & 0.461  & 0.133  & 0.307  & 0.099  \\
                                                                         & 1,0                           & 0.180  & 0.421  & 0.128  & 0.272  \\
                                                                         & 0,1                           & 0.399  & 0.122  & 0.367  & 0.112  \\
                                                                         & 1,1                           & 0.169  & 0.348  & 0.150  & 0.333  \\ \hline
\end{tabular}
\end{table}

\begin{table}[!hbp]
\centering
  \caption{The output distribution for superdense coding protocol for different values of time delay with the correction of the coherent error. For each input, 8192 runs of the algorithm on 16-qubit IBMqx5 device have been performed.}
\begin{tabular}{|c|c|c|c|c|c|}
\hline
\multirow{2}{*}{Time, $\mu s$} & \multirow{2}{*}{$a_1$, $a_2$} & \multicolumn{4}{c|}{$b_1$, $b_2$}    \\ \cline{3-6}
                          &                        & 0,0   & 1,0   & 0,1   & 1,1   \\ \hline
\multirow{4}{*}{0,0}      & 0,0                    & 0.907 & 0.039 & 0.040 & 0.013 \\
                          & 1,0                    & 0.139 & 0.801 & 0.023 & 0.036 \\
                          & 0,1                    & 0.156 & 0.027 & 0.771 & 0.046 \\
                          & 1,1                    & 0.033 & 0.119 & 0.117 & 0.731 \\ \hline
\multirow{4}{*}{0.9}      & 0,0                    & 0.862 & 0.054 & 0.056 & 0.028 \\
                          & 1,0                    & 0.150 & 0.777 & 0.033 & 0.040 \\
                          & 0,1                    & 0.147 & 0.055 & 0.722 & 0.075 \\
                          & 1,1                    & 0.051 & 0.112 & 0.130 & 0.707 \\ \hline
\multirow{4}{*}{1.8}      & 0,0                    & 0.817 & 0.069 & 0.076 & 0.039 \\
                          & 1,0                    & 0.163 & 0.737 & 0.050 & 0.051 \\
                          & 0,1                    & 0.159 & 0.085 & 0.657 & 0.099 \\
                          & 1,1                    & 0.068 & 0.125 & 0.137 & 0.670 \\ \hline
\multirow{4}{*}{2.8}      & 0,0                    & 0.760 & 0.081 & 0.102 & 0.057 \\
                          & 1,0                    & 0.169 & 0.710 & 0.063 & 0.058 \\
                          & 0,1                    & 0.181 & 0.108 & 0.602 & 0.109 \\
                          & 1,1                    & 0.084 & 0.129 & 0.144 & 0.643 \\ \hline
\multirow{4}{*}{3.7}      & 0,0                    & 0.709 & 0.092 & 0.131 & 0.068 \\
                          & 1,0                    & 0.180 & 0.674 & 0.078 & 0.068 \\
                          & 0,1                    & 0.205 & 0.119 & 0.564 & 0.111 \\
                          & 1,1                    & 0.093 & 0.140 & 0.159 & 0.608 \\ \hline
\multirow{4}{*}{4.6}      & 0,0                    & 0.656 & 0.107 & 0.160 & 0.076 \\
                          & 1,0                    & 0.181 & 0.647 & 0.088 & 0.084 \\
                          & 0,1                    & 0.215 & 0.125 & 0.541 & 0.119 \\
                          & 1,1                    & 0.110 & 0.133 & 0.156 & 0.601 \\ \hline
\end{tabular}
\end{table}



Measurements for the superdense coding protocol have been performed between April 25, 2018 and May 21, 2018.

\newpage

\section{Correction of the coherent error}

\begin{figure}[!hbp]
\includegraphics[width=.9\linewidth]{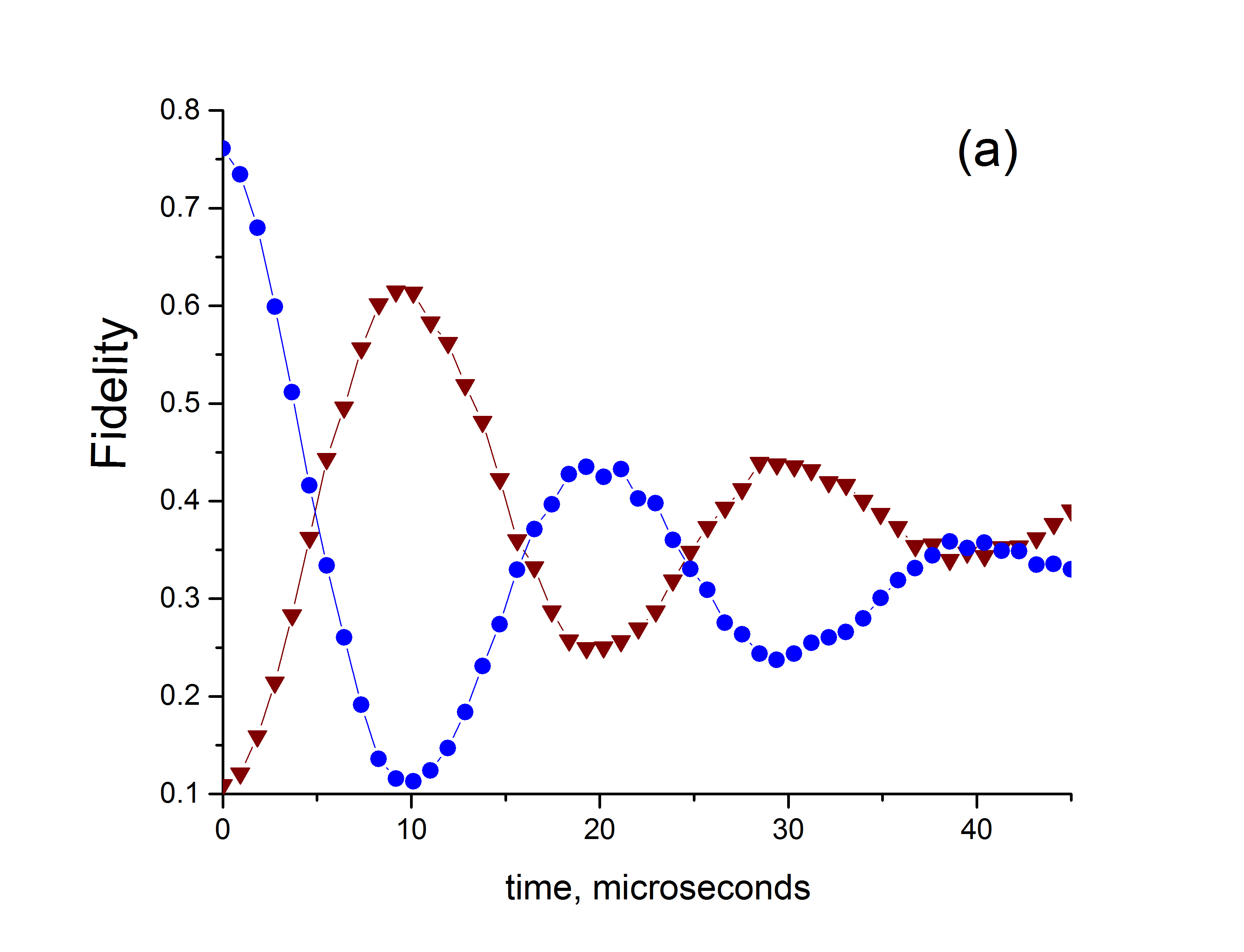}
\includegraphics[width=.9\linewidth]{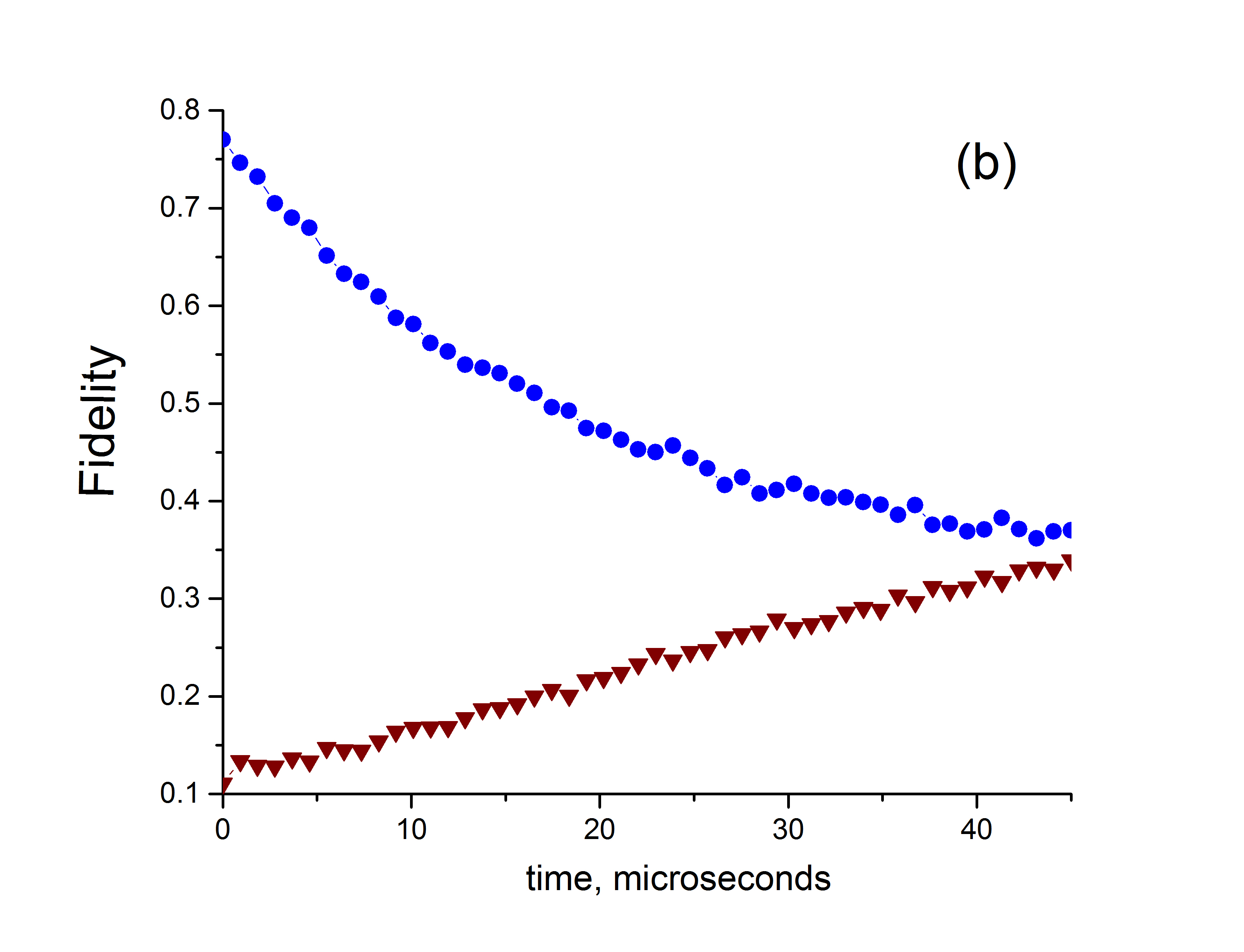}
\caption{
\label{osc}
An overlap between the prepared state and the Bell states $|\Psi_+ \rangle$ (blue circles) and $|\Psi_- \rangle$ (brown triangles) as a function of time, provided the initial target state for $|\Psi \rangle$ was  $|\Psi_+ \rangle$; (a) corresponds to direct measurements, (b) deals with the results after the correction of the coherent error (see in the text).}
\end{figure}

Fig. \ref{osc} shows the experimentally determined overlap (fidelity) between the prepared state and the Bell states $|\Psi_+ \rangle$ (blue circles) and $|\Psi_- \rangle$ (brown triangles) as a function of time, provided the initial target state was  $|\Psi_+ \rangle$. Figure \ref{osc} (a) corresponds to direct measurements, while Fig. \ref{osc} (b) deals with the results after our error correction, which compensates the drift of the internal phase. Similar oscillations have been also revealed for Bell states $|\Phi_+ \rangle$ and $|\Phi_- \rangle$.

\newpage
\section{Error distributions for BB84 protocol}

Table 5 gives error distribution for different time delays and each possible choice of the basis and bit of information, as obtained from 5-qubit IBMQx4 device. In the ideal case, the errors should absent.

\begin{table}[!hbp]
\centering
  \caption{The error distribution for BB84 protocol for different values of time delay. For each input, 8192 runs of the algorithm on 5-qubit IBMqx4 device have been performed.}
\begin{tabular}{|c|c|c|c|c|c|c|}
\hline
\multicolumn{1}{|l|}{\multirow{2}{*}{Basis, bits}} & \multicolumn{6}{c|}{Time, $\mu s$}                 \\ \cline{2-7}
\multicolumn{1}{|l|}{}                           & 0.0   & 1.2   & 2.4   & 3.6   & 4.8   & 6.0   \\ \hline
$+$,0                                              & 0.008 & 0.011 & 0.009 & 0.010 & 0.008 & 0.005 \\ \hline
$\times$,0                                              & 0.011 & 0.027 & 0.052 & 0.081 & 0.098 & 0.120 \\ \hline
$+$,1                                              & 0.051 & 0.076 & 0.095 & 0.119 & 0.177 & 0.251 \\ \hline
$\times$,1                                              & 0.050 & 0.071 & 0.091 & 0.122 & 0.176 & 0.260 \\ \hline
\end{tabular}
\end{table}

Table 6 provides error distribution for different number of SWAPs and each possible choice of the basis and bit of information, as obtained from 5-qubit IBMQx4 device. Table 7 gives similar data, but using the encoding of the logical qubit into two physical qubits supplemented by post-selection procedure. The brackets contain fraction of algorithm's runs used after the post-selection. The post-selection allowed us to improve the results, as seen from the comparison of data from Tables 6 and 7. We also note that the fraction of discarded data grows with the number of SWAPs and this leads to the improvement of the performance.

\begin{table}[!hbp]
\centering
  \caption{The error distribution for BB84 protocol for different number of SWAPs. For each input, 8192 runs of the algorithm on 5-qubit IBMqx4 device have been performed.}
\begin{tabular}{|c|c|c|c|c|}
\hline
\multirow{2}{*}{Basis, bits} & \multicolumn{4}{c|}{SWAPs}    \\ \cline{2-5}
                           & 0     & 2     & 4     & 6     \\ \hline
$+$,0                      & 0.009 & 0.036 & 0.062 & 0.078 \\ \hline
$\times$,0                 & 0.009 & 0.043 & 0.077 & 0.084 \\ \hline
$+$,1                      & 0.061 & 0.092 & 0.125 & 0.184 \\ \hline
$\times$,1                 & 0.053 & 0.089 & 0.133 & 0.175 \\ \hline
\end{tabular}
\end{table}

\begin{table}[!hbp]
\centering
  \caption{The error distribution for BB84 protocol for different number of SWAPs. Each logical qubit has been composed from two physical qubits. Post-selection procedure has been applied. For each input, 8192 runs of the algorithm on 5-qubit IBMqx4 device have been performed. Numbers in brackets indicate fractions of data accepted after the post-selection.}
\begin{tabular}{|c|c|c|c|c|}
\hline
\multirow{2}{*}{Basis, bits} & \multicolumn{4}{c|}{SWAPs}                                \\ \cline{2-5}
                           & 0            & 2            & 4            & 6            \\ \hline
$+$,0                      & 0.003 (90\%) & 0.028 (85\%) & 0.048 (79\%) & 0.076 (75\%) \\ \hline
$\times$,0                 & 0.024 (86\%) & 0.053 (84\%) & 0.081 (81\%) & 0.111 (78\%) \\ \hline
$+$,1                      & 0.002 (89\%) & 0.029 (82\%) & 0.059 (77\%) & 0.094(71\%)  \\ \hline
$\times$,1                 & 0.021 (83\%) & 0.05 (76\%)  & 0.089 (70\%) & 0.139 (63\%) \\ \hline
\end{tabular}
\end{table}

Measurements for the BB84 protocol have been performed between April 4, 2018 and May 21, 2018.

\FloatBarrier

\end{document}